\documentstyle[12pt,emlines]{article}
\input tcilatex
\begin{document}

\author{M.E. Palistrant}
\title{Upper critical field $H_{c_2}$ in two-band superconductors}
\maketitle

{\it Institutul of Appleid Physics, Chishinau, Moldova}
\\
\\
E-mail:{\it statphys@asm.md}
\\
\\

It was obtained the equation for the definition of upper critical field $%
H_{c_2}$ in the system with two overlapping energetic bands on the Fermi
surface, which is valid for the whole temperature interval $0 \leq T < T_c$.
The analytic expressions for the value $H_{c_2}$ at $T \sim T_c$ and $T \ll
T_c$ cases were defined. The possibility of changing the curvature $H_{c_2}
(T)$ with changing of the electrons speed ratio $v_1 / v_2$ on different
cavities on the Fermi surface was revealed. It was obtained the qualitative
accordance with experimental data for the intermetallic compound $MgB_2$.


\section{Introduction}

The discovery of the high temperature of superconductive transition $T \sim
40K$ in the simple intermetallic compound $MgB_2$ \cite{Nagamatsu} has
simulated researches of this material properties both in the experimental and
theoretical plans. An overview of the basic physical properties of $MgB_2$
can be seen, for example, in \cite{Canfield}. The significant result of
these researches is the discovery of two energetic gaps in the spectrum of the
elementary excitations \cite{Chen}, \cite{Tsuda} and the possibility of
theoretical describing of this compound on the base of the two-band model
\cite{Moskalenko} (see also \cite{Suhle}).

The theory of the thermodynamic and kinematic properties of superconductors
(pure and doped) with overlapping energy bands on the Fermi surface was
developed by Moskalenko and his collaborators (for the references to the
articles see \cite{Palistrant} - \cite{Kon}). The typical feature of
two-band model is the fact, that is gives not only the quantitative difference
from the one energy band case, but leads to the qualitative new results
\cite{Palistrant}.  Being built long before the discovery of the
superconductivity in $MgB_2$, this theory can generally describe the
superconductive properties of this compound.  Nowadays often happens
rediscovering of already known results, which had been obtained in the
previously mentioned articles.

Moskalenko two-band model \cite{Moskalenko} and its generalization for the
anisotropic value of the energetic gaps $\Delta_1$ and $\Delta_2$ case
\cite{Mishonov}, \cite{Mishonov_1},  confirm the experimental results
for the thermal capacity $C_S$, dependence on temperature, the peneration dept
of the magnetic field $\lambda (T)$ and other characteristics in the
superconducting $MgB_2$ compound.

At it is known, the superconductive metals undergo the transition from the
superconductive phase into normal in the magnetic field at some its value.
Researching the Ginzburg-Landau equation, Abrikosov has shown
\cite{Abrikosov} that at some value of the exterior magnetic field $H_{c_1}$
(lower critical field) the mixed state realizes in the secondary type
superconductors. This state is characterized by penetration of the magnetic
field in the deep of superconductor as the exact lattice of the magnetic lines.
The transitions into the normal state, which is relevant to the full
penetration of the magnetic field into the superconductor, occurs in the moment
when the field achieves the value of the upper critical field $H_{c_2}$.
Definition of this value on the base of the Ginzburg-Landau theory is possible
only in the critical temperature neighbourhood. The $H_{c_2}$ value is defined
on the whole temperature interval in Goricov article \cite{Gor'kov} (see
also Maki and Tsuzuki \cite{Maki} ).

In the previously mentioned articles was considered an isotropic
superconductor. It is interesting to research the magnetic properties of
anisotropic superconductors because so are real superconductors. Overlapping of
the energy bands \cite{Moskalenko} is one of the manifestation of
anisotropy.  There were obtained base equations of the electrodynamics of
two-band superconductors in the work \cite{Moskalenko_2}, which are valid
both for pure and doped superconductors.  Also were researched the
electromagnetic properties of two-band superconductors for temperatures close
to the critical $(T_c - T \ll T_c)$.

The main purpose of the work is researching of pure
two-band superconductor of the secondary type for arbitrary temperatures and
external magnetic field close to the upper critical field and the definition
of temperature dependence of the $H_{c_2}$ value.

\section{The system of equations for the order parameters $\Delta_n$.}

If the exterior magnetic field is great enough,the order parameters
$\Delta_m\,(m = 1,2)$of two-band superconductor is  small enough, and we can
use equations ref. \cite{Moskalenko_2} for pure two-band superconductor:
\begin{equation}
\Delta_{m}^{*} (\vec x) = \frac{1}{\beta}\sum_{\omega}\,\sum_{n n'} V_{n m}
\int\,d{\vec y} g_{n'n} (\vec y, \vec x / \omega) \Delta_{n'}^{*}(\vec y)
g_{n'n} (\vec y, \vec x/-\omega).
\end{equation}

We restricted here  by linear terms on $\Delta_n$ quantities in comparison with
given in \cite{Moskalenko_2} because in the $H = H_{c_2}$ point occurs
solutions with the infinitely small values $\Delta_m$. Green function defines
by equation at presence of the magnetic field  \cite{Gor'kov_1}:
\begin{equation}
g_{nn'}(r, r'/\omega) = e^{i \varphi (r, r')} g_{n'n}^{0}\,(r,r'/\omega)
\end{equation}

Were $g_{n'n}^{0}$ - Green function of an electron in normal
metal without magnetic field. The presence of the magnetic field is taken into
account by the phase multiplier
\begin{equation}
\varphi (r,r')=e \int \limits_{r'}^{r}\,A(\vec l) d \vec l.
\end{equation}

We decompose in equation (1) the normal metal function $g_{n'n}^{0}$ into the
row by he Bloch functions $\psi_{n \vec k} (\vec x) = e^{i(\vec k \vec x)} U_{n
\vec k} (\vec x)/ \sqrt N$ ($U_{n \vec k}$ - the Bloch amplitude):
\begin{equation}
g_{n'n}^{0}(\vec y, \vec x/\omega) = \sum_{\vec k \vec k'} g_{n'n}^{0} (\vec
k', \vec k /\omega)\psi_{n'\,\vec k'} (\vec y) \psi_{n \vec k}^{*} (\vec x)
\end{equation}

and use approaching of the diagonal
Green functions, which were developed in works \cite{Moskalenko_3}.
Besides choose the vector potential as
\begin{equation}
A_x = A_z = 0 ;\,\,\,\,\,A_y = H_0 x. 
\end{equation}
(magnetic field is guided along the z axis).

Equation (1) will accept the next view:
$$
\Delta_{m}^{*}(r) = \sum_n V_{nm}\frac{1}{\beta}\sum_{\omega}\int d^3 r'
\Delta_{n}^{*}(r') \int \frac{d^3 k d^3 q}{(2 \pi)^6}g_{n}^{0}(\vec k/\omega)
\times
$$
$$
\times g_{n}^{0}\left(\vec q - \vec k /-\omega\right)\,exp \left[i \vec q(\vec
r'- \vec r) + ie(x+x') (y'- y) H_0 \right]\,U_{n \vec k}(\vec r') \times
$$
\begin{equation}
\times U_{n \vec k}^{*}(\vec r) U_{n \vec q - \vec k}(r') U_{n \vec q - \vec
k}(r),
\end{equation}

\begin{equation}
g_{n}^{0}(\vec k /\omega) = \frac{1}{i\omega - \xi (k)}.
\end{equation}

After integrating by impulse $k$,
we make the averaging by elementary cells, in order to exclude from
consideration fast oscillations of function $\Delta_n$ (which are stipulated
by Bloch functions) and to save just dependence on these functions' coordinates
(incipient due to the exterior magnetic field presence), because just last
dependence is observed.  As a result we obtain:
$$
\Delta_{m}^{*}(r) = \sum_n V_{nm} \frac{1}{\beta} \sum_{\omega}\int d \Omega
\int \frac{d^3 q}{(2 \pi)^3} \frac{N_n \omega}{2 |\omega|} \frac{i}{2 i \omega
+ \vec q \vec v_n}\times
$$
\begin{equation}
\times \int d^3 r' \Delta_{n}^{*}(r') exp \left[iq (r - r') + ie H_0 (x + x')
(y' - y)\right],
\end{equation}
where $v_n$ and $N_n$ - are  accordingly electron speed and electron density
of state on n-th cavity of the Fermi surface.  Applying to the calculation of
the equation's (8) left part Maki and Tsuzuki methodic \cite{Maki}, and
supposing that $\Delta_n$ depends just on $x^|$, we obtain on this way:
$$
\Delta_{m}^{*}(x) = \sum_n \frac{V_{nm} \pi N_n}{\beta v_n} \int \limits_{1}^
{\infty} \frac{d u}{u} \int \limits_{-\infty}^{\infty} d x'\Delta_{n}^{*}(x')
\times
$$
\begin{equation}
\times \frac{J_0 \left[(x^2 - x'^2)e H_0 \sqrt{u^2 - 1}\right]}{sh \frac{2
|x - x'| \pi u}{\beta v_n}} - \theta \left(|x-x'| - \delta_n\right).
\end{equation}
The introduction of $\delta_n$  corresponds to cutting of the inter-electron
interaction in the impulse space in vicinity of n-th cavity of the Fermi
surface. It is easy to notice, that at $T = T_c$, $H_{c_2} = 0$ equation (9)
turns to the equation for the critical temperature definition.  On the base of
the last equation the $\delta_n$ value defines:
$$
\delta_n = \frac{v_n}{2 \gamma e_0 \omega_D},
$$
$e_0$ - is the base of natural logarithm, $\omega_D$ - Debay freguency. The
solution of the equation (9) we search in the view \cite{Moskalenko_3}:
\begin{equation}
\Delta_m (x) = \Delta_{m} e^{- e H_0 x^2}. 
\end{equation}

Using formula
(10), we can obtain equation (9) in the next view:
$$
\Delta_{m}^{*} = \sum_n \sqrt{\frac{2}{\pi}}\,\frac{V_{nm}\pi N_n}{\beta v_n}
\Delta_{n}^{*}\int \limits_{1}^{\infty} \frac{d u}{u} \int \limits_{-\infty}^
{\infty} d x \int \limits_{-\infty}^{\infty}d x'\theta \left(|x - x'|-
\delta_n \right) \times
$$
$$
\times exp \left[- e H_0 (x^2 +x'^2)\right] \frac{J_0 \left[(x^2 - x'^2) e
H_0 \sqrt{u^2 - 1}\,\right]}{sh \frac{2 |x - x'| \pi u}{\beta v_n}}=
$$
\begin{equation}
= \sum_n V_{nm}N_n \rho_{n}^{-1/2}\Delta_{n}^{*} \int \limits_{1}^{\infty}
\frac{du}{u} \int \limits_{\delta_{n}^{'}}^{\infty} d \zeta exp
\left[-\frac{\zeta^2} {4}(u^2 + 1)\right] \frac{I_0 \left[\frac{\zeta^2}{4}(u^2
- 1) \right]}{sh \left[u \zeta \rho_{n}^{-1/2}\right]},
\end{equation}
were
\begin{equation}
\delta'_{n} = (e H_0)^{1/2} \delta_n;\,\,\,\,\,\,\,\,\,\,\,\,\,\,\rho_n =\frac
{v_{n}^{2} e H_0}{(2 \pi T)^2}.
\end{equation}

\section{Upper critical field $H_{c_2}$ definition.}

Basing on (11) it is easy to obtain the
equation for the upper critical field definition in the next view:
\begin{equation}
-\Delta_{m}^{*} + \sum_n V_{nm}\,N_n \Delta_{n}^{*} ln \frac{2 \gamma
\omega_{D}^{n}}{\pi T_c}+ \sum_n V_{n m} N_n \Delta_{n}^{*}\left[ln\frac{T_c}
{T} - f (\rho_n)\right] = 0,
\end{equation}
where
$$
f(\rho_n) = \rho_{n}^{-1/2}\int \limits_{0}^{\infty} d \zeta \int
\limits_{1}^{\infty} \frac{du}{u\,sh [u \zeta \rho_{n}^{-1/2}]}\times
$$
\begin{equation}
\times \Biggl\{1 - exp \left[-\frac{\zeta^2}{4}(u^2 + 1)\right] I_0
\left[\frac{\zeta^2}{4}(u^2 - 1)\right] \Biggr\}.
\end{equation}
Here was used the next correlation:
\begin{equation}
\rho_{n}^{1/2}\,\int \limits_{1}^{\infty}  \frac{du}{u}  \int
\limits_{\delta_{n}^{'}}^{\infty} d \zeta \frac{1}{sh [ u \zeta
\rho_{n}^{-1/2}]} = ln \frac{2 \rho^{1/2}}{\delta_{n}^{'} e_0},
\end{equation}
Equality to zero of the system (13) determinant corresponds to the presence of
non-zero solutions, that is the connected pairs forming. The field, in the
presence of which such solutions can appear, is the upper critical field
$H_{c_2}$. So the $H_{c_2}$ value defines from the condition of the system (13)
solvability:
\begin{equation}
a f (\rho_1) f(\rho_2)+B_1 f(\rho_1) + B_2 f(\rho_2)+ C = 0,
\end{equation}
where
$$
B_n = N_n V_{nn}- a \xi_{c}^{(n)};\,\,\,\,\,\,\,\,\,\,\,\,(n = 1,2)
$$
$$
C = 1 - N_1 V_{11} \xi_{T}^{(1)} - N_2 V_{22} \xi_{T}^{(2)} + a \xi_{T}^{(1)}
\xi_{T}^{(2)};\,a = N_1 N_2 (V_{11}V_{22} - V_{12}V_{21});
$$

$$
\xi_{T}^{(n)} = ln \frac{2 \gamma \omega_{D}^{(n)}}{\pi T};\,\,\,\,\,\,
\xi_{c}^{(n)} = ln \frac{2 \gamma \omega_{D}^{(n)}}{\pi T_c}
$$
Condition (16) considers in the two limit cases:\\
\\
a. $\rho_n \ll 1 (T_c - T \ll T_c);\,\,\,\,\,$b. $\rho_n \gg 1 (T \ll T_c)$,\\
\\
for which functions $f(\rho_n)$ are defined in works \cite{Maki} :
\begin{equation}
f(\rho_n)= \frac{7}{6} \zeta(3) \rho_n -\frac{31}{10} \zeta(5) \rho_{n}^{2} +
\frac{381}{28} \zeta(7) \rho_{n}^{3} ,\,\,\,\,\,\,\,\rho_n \ll 1.
\end{equation}
\begin{equation}
f(\rho_n) = ln \frac{2 (2 \gamma \rho_n)^{1/2}}{e_0}- \frac{1}{\pi^2 \rho_n}
\left[\zeta'(2) + \frac{\zeta(2)}{2} ln\frac{2}{\pi^2 \gamma
\rho_n}\right],\,\,\,\rho_n \gg 1.
\end{equation}
Case a.$\rho_n \ll 1$\,\,\,  $(T_c - T \ll T_c)$
Subject  to the low electronic-phonon binding, expression (16) accepts the next
view in this case:
$$
B_1 f(\rho_1) + B_2 f(\rho_2) + C_a = 0 ,
$$
were
\begin{equation}
C_a = \left[-(N_1 V_{11} + N_2 V_{22}) + 2a \xi\right] ln\frac{T_c}{T}.
\end{equation}
\\
(It is supposed as usually $\xi_{c}^{(1)} = \xi_{c}^{(2)} = \xi_c = ln
\frac{2 \omega_D \gamma}{\pi T_c}$).\\
During obtaining of (19) had been used
equality to zero determinant of the next system of equations:
\begin{equation}
\Delta_{m c}^{*} = \sum_n V_{nm} N_n \Delta_{nc}^{(*)} \xi_c,
\end{equation}
This system defines the critical temperature Tc of superconductor. Basing
on formulas (19) and (20), (17) and (12) we obtain expression for the $H_{c_2}$
value (see also \cite{Palistrant_1}, \cite{Moskalenko_4}):

$$
H_{c_2}(T) = \frac{4 \pi^2 T_{c}^{2}}{e}\left[v_{1}^{2}\eta_1+ v_{2}^{2}\eta_2
\right]^{-1} \frac{6}{7 \xi (3)}\theta.
$$
\begin{equation}
\left[1 + \theta \Biggl\{\frac{\frac{v_{1}^{2}}{v_{2}^{2}}\eta_1 +
\frac{v_{2}^{2}}{v_{1}^{2}}\eta_2}{\left(\frac{v_1}{v_2}\eta_1 +
\frac{v_2}{v_1}\eta_2\right)}\,\frac{31}{10} \xi (5) \left(\frac{6}{7 \xi (3)}
\right)^2 - \frac{3}{2}\Biggr\}\right],
\end{equation}
$$
\theta = 1 - \frac{T}{T_c},\,\,\,\eta_1 = \frac{1}{2}(1 + \eta)\,\,\,\eta_2 =
\frac{1}{2}(1 - \eta);
$$
$$
\eta = \frac{N_1 V_{11} - N_2 V_{22}}{\sqrt{(N_1 V_{11} - N_2 V_{22})^2 +
4N_1 N_2 V_{12} V_{21}}}.
$$
Coefficients $\eta_1$, $\eta_2$, $\eta$ satisfy the expressions:
$$
0 \leq \eta_1 \leq 1,\,0 \leq \eta_2 \leq 1,\,\eta_1 + \eta_2 = 1,\, 0 \leq
|\eta| \leq 1.
$$
Case b. $\rho_n \gg 1$\,\,\,\,\, $(T \ll T_c)$

Basing on equation (16) and using (12) and
(18) we obtain for the $H_{c_2}(T)$ value the next expression in this case:
\begin{equation}
H_{c_2} (T) = \frac{2 \gamma\, \omega_{D}^{2}}{e v_1 v_2} exp (2 - \Sigma),
\end{equation}
were
\begin{equation}
\Sigma = \eta^+ \pm \sqrt{(ln \lambda - \eta^{(-)})^2 + \frac{4 N_1 N_2 V_{12}
V_{21}}{a^2} - 4 F(\rho_0)},
\end{equation}
$$
\eta^{\pm} = \frac{1}{a}\left(N_1 V_{11} \pm N_2
V_{22}\right),\,\,\,\,\,\lambda = \frac{v_1}{v_2},
$$
\begin{equation}
F(\rho_0) = \frac{1}{\rho_0} \left[P(ln\, \rho_0)^2 + Q ln\, \rho_0 +
R\right], 
\end{equation}

\begin{equation}
\rho_0 = \frac{e v_1\,v_2}{4 \pi^2 T^2}\,H_{c_2}(0).
\end{equation}

\begin{equation}
H_{c_2}(0) = \frac{2 \gamma \omega_{D}^{2}}{e v_1 v_2}exp(2 - \alpha),
\end{equation}

\begin{equation}
\alpha = \eta^{(+)} \pm \sqrt{(ln \lambda - \eta^{(-)})^2 + \frac{4 N_1 N_2
V_{12} V_{21}}{a^2}}
\end{equation}

\begin{equation}
P = \frac{\zeta (2)}{4\pi^2} \left(\lambda + \frac{1}{\lambda}\right),
\end{equation}

$$
Q = \frac{1}{2 \pi^2}\left(\lambda + \frac{1}{\lambda}\right)
\left[\frac{\zeta (2)}{2}\,ln (x_1 x_2) - \zeta' (2)\right] +
$$
\begin{equation}
+ \frac{\zeta (2)}{2 \pi^2} \left[\varepsilon^{(+)} - \left(\lambda +
\frac{1}{\lambda}\right) \xi_T\right],\,\,x_1 = \frac{8 \gamma}{e_{0}^{2}},
x_2 = \frac{\pi^2 \gamma}{2};
\end{equation}

$$
R = \frac{\zeta (2)}{4 \pi^2}\Biggl\{ \left(\lambda +
\frac{1}{\lambda}\right)\left[ ln\,x_1\,ln\,x_2 - (ln\, \lambda)^2\right] -
\left(\lambda - \frac{1}{\lambda}\right) ln\, \lambda\, ln
\frac{x_1}{x_2}\Biggr\}-
$$
$$
- \frac{\xi' (2)}{2 \pi^2}\left[\left(\lambda + \frac{1}{\lambda}\right)
ln\,x_1 + \left(\lambda - \frac{1}{\lambda}\right)\,ln\,\lambda\right] +
\frac{1}{\pi^2}\left[\frac{\xi(2)}{2}\,ln x_2 - \zeta' (2)\right]\times
$$
$$
\times \left[\varepsilon^{(+)} - \left(\lambda + \frac{1}{\lambda}\right)
\xi_T \right] - \frac{\zeta (2)}{2 \pi^2}\,ln \lambda \left[\varepsilon^{(-)}-
\left(\lambda - \frac{1}{\lambda}\right) \xi_T\right],
$$
\begin{equation}
\varepsilon^{(\pm)} = \frac{1}{\alpha}\left(\lambda N_2 V_{22} \pm
\frac{1}{\lambda} N_1 V_{11}\right).
\end{equation}
In our case (T close to zero) expression can be adduced to the view:
\begin{equation}
H_{c_2}(T) = H_{c_2} (0) \left[1 + \frac{2}{s v (\lambda)} F (\rho_0)\right]
\end{equation}
were
\begin{equation}
\nu(\lambda) = \sqrt{(ln \lambda - \eta^{(-)})^2 + \frac{4 N_1 N_2 V_{12}
V_{21}}{a^2}},\,\,\,\,\,\,\,\,\,\,s = \pm 1
\end{equation}
or
$$
\frac{H_{c_2}(T)}{H_{c_2}(0)}= 1 + \frac{16 \gamma}{e_{0}^{2}
\pi^2}\left(\frac{T}{T_c}\right)^2 e^{s \nu (\lambda) -
s \nu(1)}\left\{\left(\lambda \gamma^+ + \frac{1}{\lambda}\gamma^-\right)\right.
$$
$$
\left[\zeta' (2) + \zeta(2) ln \frac{4 T}{e_0 \pi T_c}+\frac{s}{2} \nu
(\lambda) - \frac{s}{2} \nu(1)\right]+
$$
\begin{equation}
\left.+\frac{\zeta (2)}{2}\left(\lambda
\gamma^+ - \frac{1}{\lambda} \gamma^- \right)ln\lambda \right\},
\end{equation}

\begin{equation}
\gamma \pm = \frac{1}{2}\left[1 \pm \frac{\eta^{(-)} - ln \lambda}{s \nu
(\lambda)}\right],\,\,H_{c_2} (0) = \frac{\pi^2\,T_{c}^{2} e_{0}^{2}}{2
\gamma e \nu_1\,\nu_2} exp \left[\nu (1) - \nu (\lambda)\right].
\end{equation}
Expression (33) can be also obtained from formula (33) of work
\cite{Palistrant_1}.  Let's consider simplified case $V_{22} = 0$. At that
on the base of definitions (24) and (32) we obtain:
$$
\eta^{(\pm)} = -\frac{\lambda_{11}}{\lambda_{12}
\lambda_{21}},\,\,\,\lambda_{11} = N_1 V_{11}\,\,\,\lambda_{12} = N_2 V_{12},
\,\,\,\lambda_{21} = N_1 V_{21}
$$
$$ \nu(\lambda) = \sqrt{\left(ln \lambda +
\frac{\lambda_{11}}{\lambda_{21}\lambda_{12}}\right)^2 +
\frac{4}{\lambda_{12}\,\lambda_{21}}}\,\,\,\,,
$$
\begin{equation}
\gamma^{\pm} = \frac{1}{2}\left[1 \mp \frac{\frac{\lambda_{11}}{\lambda_{12}
\lambda_{21}}+ ln \lambda}{s \nu(\lambda)}\right]
\end{equation}
For $s = 1$ and $V_{22}=0$ the expression for $H_{c_2}(T)$ (33) turns into
corresponding expression of work  \cite{Moskalenko_1}. For the temperature
hear to the value $T_c$ in the limit case $V_{22} = 0$ we obtain:
$$
\frac{H_{c_2}(T \sim T_c)}{H_{c_2}(0)} = \frac{8 \gamma v_1\,v_2}{e_{0}^{2}
\left[v_{1}^{2} \eta_1 + v_{2}^{2} \eta_2 \right]} exp (\nu(\lambda) - \nu(1))
\frac{6}{7 \xi (3)}\,\,\times
$$
\begin{equation}
\times \left(1 - \frac{T}{T_c} \right) \Biggl\{1+ \left(1 - \frac{T}{T_c}
\right) \left[\frac{\frac{v_{1}^{2}}{v_{2}^{2}}\eta_1 +
\frac{v_{2}^{2}}{v_{1}^{2}}\eta_2}{\left(\frac{v_1}{v_2} \eta_1 +
\frac{v_2}{v_1}\eta_2\right)^2} \frac{31}{10} \zeta(5) \left(\frac{6}{7 \zeta
(3)}\right)^2 - \frac{3}{2}\right]\Biggr\},
\end{equation}
were
\begin{equation}
\eta_{1,2} = (1 \pm \eta)/2,\,\,\,\,\,\,\,\,\,\,\,\,\,\,\,\eta =
\frac{\lambda_{11}}{\lambda_{12} \lambda_{21} \nu(1)}
\end{equation}
Assuming in  formulas (33)-(36) $v_1 = v_2$ we obtain corresponding ratios for
usual one-band superconductor. These ratios have the view:
\begin{equation}
\frac{H_{c_2} (T \rightarrow 0)}{H_{c_2}(0)} = 1 + \frac{16 \gamma}{\pi^2
e_{0}^{2}}\left(\frac{T}{T_c}\right)^2 \Biggl\{\zeta (2) ln \frac{T}{T_c} +
\zeta'(2) + \zeta(2) ln \frac{4}{\pi e_0}\Biggr\},
\end{equation}
\begin{equation}
\frac{H_{c_2} (T \rightarrow T_c)}{H_{c_2}(0)} = \frac{8 \gamma}{e_{0}^{2}}\,\,
\frac{6}{7 \zeta (3)}\left(1 - \frac{T}{T_c}\right)\Biggl\{1 + \left(1 -
\frac{T}{T_c}\right)\left[\frac{31}{10}\zeta(5)\left(\frac{6}{7 \zeta
(3)}\right)^2\,- \frac{3}{2}\right]\Biggr\}
\end{equation}
and coincide with the results of works \cite{Gor'kov}, \cite{Maki}.

\section{The discussion of the results.}

We obtained  equation (16), on the base of which the value of the upper
critical field in the two-band system can be calculated on the whole
temperature interval $0 \leq T < Tc$. The analytic solutions of this
equation were obtained for $T \rightarrow T_c$ (21) and $T\rightarrow 0$ (33 ).
It is easy to notice that $H_{c_2}$ depends on the correlations of the speeds
$v_1$ and $v_2$ of the electrons on the Fermi surface, and on the constants of
the electronic-phonon interaction $\lambda_{nm}$.

If $H_{c_2}^{0}$ and $T_{c_0}$ are introduced (upper critical field and
critical temperature of the one-band low-temperature superconductor), on the
base of (34) we obtain:
\begin{equation}
H_{c_2}(0)/H_{c_2}^{0}(0) = (T_c/T_{c_0})^2 \frac{v_1}{v_2} exp((v(1)
- v(\lambda))
\end{equation}
The numerical estimations (40) let us do the  conclusion, that the upper
critical field of two-band superconductors for $T = 0$ can exceed the value of
$H_{c_2}^{0}(0)$ for usual superconductors by two-three orders.  These big
values $H_{c_2}(0)$ are provided by high $T_c$ and by ratio $v_1/v_2 > 1$
or $\gg 1$.

The dependence  $H_{c_2}(T)/H_{c_2}(0)$, which was obtained on the base of
formulas (33) and (34) of the value $H_{c_2}$ for $T \sim 0$ and $T \sim T_c$
correspondingly and their extrapolations, are given in the fig.1.
\\
\\
\special{em:linewidth 0.4pt}
\unitlength 0.8mm
\linethickness{0.4pt}
\begin{picture}(132.67,119.16)
\emline{32.33}{18.67}{1}{32.33}{119.00}{2}
\emline{32.33}{18.67}{3}{132.66}{18.67}{4}
\emline{42.66}{18.67}{5}{42.66}{21.34}{6}
\emline{52.66}{21.34}{7}{52.66}{18.67}{8}
\emline{62.66}{18.67}{9}{62.66}{21.00}{10}
\emline{72.66}{21.00}{11}{72.66}{18.67}{12}
\emline{82.66}{18.67}{13}{82.66}{21.00}{14}
\emline{92.66}{21.00}{15}{92.66}{18.67}{16}
\emline{102.66}{18.67}{17}{102.66}{21.00}{18}
\emline{112.66}{21.00}{19}{112.66}{18.67}{20}
\emline{122.66}{18.67}{21}{122.66}{21.00}{22}
\emline{132.66}{21.00}{23}{132.66}{18.67}{24}
\emline{32.33}{29.00}{25}{32.33}{29.00}{26}
\emline{32.33}{29.00}{27}{32.33}{29.00}{28}
\emline{32.33}{29.00}{29}{32.33}{29.00}{30}
\emline{32.33}{29.00}{31}{32.33}{29.00}{32}
\emline{32.33}{29.00}{33}{32.33}{29.00}{34}
\emline{32.33}{29.00}{35}{32.33}{29.00}{36}
\emline{32.33}{29.00}{37}{32.33}{29.00}{38}
\emline{32.33}{29.00}{39}{32.33}{29.00}{40}
\emline{32.33}{29.00}{41}{32.33}{29.00}{42}
\emline{32.33}{29.00}{43}{32.33}{29.00}{44}
\emline{32.33}{29.00}{45}{32.33}{29.00}{46}
\emline{32.33}{29.00}{47}{32.33}{29.00}{48}
\emline{32.33}{29.00}{49}{32.33}{29.00}{50}
\emline{32.33}{29.00}{51}{32.33}{29.00}{52}
\emline{32.33}{29.00}{53}{32.33}{29.00}{54}
\emline{32.33}{29.00}{55}{32.33}{29.00}{56}
\emline{32.33}{29.00}{57}{32.33}{29.00}{58}
\emline{32.33}{29.00}{59}{32.33}{29.00}{60}
\emline{32.33}{29.00}{61}{32.33}{29.00}{62}
\emline{32.33}{29.00}{63}{32.33}{29.00}{64}
\emline{32.33}{29.00}{65}{32.33}{29.00}{66}
\emline{32.33}{29.00}{67}{32.33}{29.00}{68}
\emline{32.33}{29.00}{69}{32.33}{29.00}{70}
\emline{32.33}{29.00}{71}{32.33}{29.00}{72}
\emline{32.33}{29.00}{73}{32.33}{29.00}{74}
\emline{32.33}{29.00}{75}{32.33}{29.00}{76}
\emline{32.33}{29.00}{77}{32.33}{29.00}{78}
\emline{32.33}{29.00}{79}{32.33}{29.00}{80}
\emline{32.33}{29.00}{81}{32.33}{29.00}{82}
\emline{32.33}{29.00}{83}{32.33}{29.00}{84}
\emline{32.33}{29.00}{85}{32.33}{29.00}{86}
\emline{32.33}{29.00}{87}{32.33}{29.00}{88}
\emline{32.33}{29.00}{89}{32.33}{29.00}{90}
\emline{32.33}{29.00}{91}{32.33}{29.00}{92}
\emline{32.33}{29.00}{93}{32.33}{29.00}{94}
\emline{32.33}{29.00}{95}{32.33}{29.00}{96}
\emline{32.33}{29.00}{97}{32.33}{29.00}{98}
\emline{32.33}{29.00}{99}{32.33}{29.00}{100}
\emline{32.33}{29.00}{101}{32.33}{29.00}{102}
\emline{32.33}{29.00}{103}{32.33}{29.00}{104}
\emline{32.33}{29.00}{105}{32.33}{29.00}{106}
\emline{32.33}{29.00}{107}{32.33}{29.00}{108}
\emline{32.33}{29.00}{109}{32.33}{29.00}{110}
\emline{32.33}{29.00}{111}{32.33}{29.00}{112}
\emline{32.33}{29.00}{113}{32.33}{29.00}{114}
\emline{32.33}{29.00}{115}{32.33}{29.00}{116}
\emline{32.33}{29.00}{117}{32.33}{29.00}{118}
\emline{32.33}{29.00}{119}{32.33}{29.00}{120}
\emline{32.33}{29.00}{121}{32.33}{29.00}{122}
\emline{32.33}{29.00}{123}{32.33}{29.00}{124}
\emline{32.33}{29.00}{125}{32.33}{29.00}{126}
\emline{32.33}{29.00}{127}{32.33}{29.00}{128}
\emline{32.33}{29.00}{129}{32.33}{29.00}{130}
\emline{35.00}{29.00}{131}{32.33}{29.00}{132}
\emline{32.33}{39.00}{133}{35.00}{39.00}{134}
\emline{35.00}{49.00}{135}{32.33}{49.00}{136}
\emline{32.33}{59.00}{137}{35.00}{59.00}{138}
\emline{35.00}{59.00}{139}{35.00}{59.00}{140}
\emline{35.00}{59.00}{141}{35.00}{59.00}{142}
\emline{35.00}{69.00}{143}{32.33}{69.00}{144}
\emline{32.33}{79.00}{145}{35.00}{79.00}{146}
\emline{35.00}{89.00}{147}{32.33}{89.00}{148}
\emline{32.33}{99.00}{149}{35.00}{99.00}{150}
\emline{35.00}{109.00}{151}{32.33}{109.00}{152}
\emline{32.33}{119.00}{153}{35.00}{119.00}{154}
\emline{35.00}{119.00}{155}{35.00}{119.00}{156}
\emline{35.00}{119.00}{157}{35.00}{119.00}{158}
\emline{35.00}{119.00}{159}{35.00}{119.00}{160}
\emline{35.00}{119.00}{161}{35.00}{119.00}{162}
\emline{35.00}{119.00}{163}{35.00}{119.00}{164}
\emline{35.00}{119.00}{165}{35.00}{119.00}{166}
\emline{35.00}{119.00}{167}{35.00}{119.00}{168}
\emline{35.00}{119.00}{169}{35.00}{119.00}{170}
\emline{35.00}{119.00}{171}{35.00}{119.00}{172}
\emline{35.00}{119.00}{173}{35.00}{119.00}{174}
\emline{35.00}{119.00}{175}{35.00}{119.00}{176}
\emline{34.00}{114.00}{177}{32.33}{114.00}{178}
\emline{32.33}{104.00}{179}{34.00}{104.00}{180}
\emline{34.00}{104.00}{181}{34.00}{104.00}{182}
\emline{34.00}{104.00}{183}{34.00}{104.00}{184}
\emline{34.00}{104.00}{185}{34.00}{104.00}{186}
\emline{34.00}{104.00}{187}{34.00}{104.00}{188}
\emline{34.00}{104.00}{189}{34.00}{104.00}{190}
\emline{34.00}{104.00}{191}{34.00}{104.00}{192}
\emline{34.00}{104.00}{193}{34.00}{104.00}{194}
\emline{34.00}{104.00}{195}{34.00}{104.00}{196}
\emline{34.00}{104.00}{197}{34.00}{104.00}{198}
\emline{34.00}{104.00}{199}{34.00}{104.00}{200}
\emline{34.00}{104.00}{201}{34.00}{104.00}{202}
\emline{34.00}{104.00}{203}{34.00}{104.00}{204}
\emline{34.00}{104.00}{205}{34.00}{104.00}{206}
\emline{34.00}{94.00}{207}{32.33}{94.00}{208}
\emline{32.33}{84.00}{209}{34.00}{84.00}{210}
\emline{34.00}{74.00}{211}{32.33}{74.00}{212}
\emline{32.33}{64.00}{213}{34.00}{64.00}{214}
\emline{34.00}{54.00}{215}{32.33}{54.00}{216}
\emline{32.33}{44.00}{217}{34.00}{44.00}{218}
\emline{34.00}{34.00}{219}{32.33}{34.00}{220}
\emline{32.33}{24.00}{221}{34.00}{24.00}{222}
\emline{37.66}{18.67}{223}{37.66}{20.34}{224}
\emline{37.66}{20.34}{225}{37.66}{20.34}{226}
\emline{37.66}{20.34}{227}{37.66}{20.34}{228}
\emline{47.66}{20.34}{229}{47.66}{18.67}{230}
\emline{57.66}{18.67}{231}{57.66}{20.34}{232}
\emline{67.66}{20.34}{233}{67.66}{18.67}{234}
\emline{77.66}{18.67}{235}{77.66}{20.34}{236}
\emline{87.66}{20.34}{237}{87.66}{18.67}{238}
\emline{97.66}{18.67}{239}{97.66}{20.34}{240}
\emline{107.66}{20.34}{241}{107.66}{18.67}{242}
\emline{117.66}{18.67}{243}{117.66}{20.34}{244}
\emline{127.66}{20.34}{245}{127.66}{18.67}{246}
\put(27.66,18.67){\makebox(0,0)[cc]{$0$}}
\put(27.66,39.00){\makebox(0,0)[cc]{$0,2$}}
\put(27.66,59.00){\makebox(0,0)[cc]{$0,4$}}
\put(27.66,79.00){\makebox(0,0)[cc]{$0,6$}}
\put(27.66,99.00){\makebox(0,0)[cc]{$0,8$}}
\put(27.66,119.00){\makebox(0,0)[cc]{$1,0$}}
\put(52.66,13.67){\makebox(0,0)[cc]{$0,2$}}
\put(72.66,13.67){\makebox(0,0)[cc]{$0,4$}}
\put(92.66,13.67){\makebox(0,0)[cc]{$0,6$}}
\put(112.66,13.67){\makebox(0,0)[cc]{$0,8$}}
\put(132.66,13.67){\makebox(0,0)[cc]{$1,0$}}
\put(115.66,7.67){\makebox(0,0)[cc]{$T/T_c$}}
\put(10.67,106.00){\makebox(0,0)[cc]{$H_{c_2}(T)/H_{c_2}(0)$}}
\emline{33.27}{119.16}{247}{34.62}{117.94}{248}
\emline{34.62}{117.94}{249}{35.99}{116.60}{250}
\emline{35.99}{116.60}{251}{37.37}{115.15}{252}
\emline{37.37}{115.15}{253}{38.77}{113.57}{254}
\emline{38.77}{113.57}{255}{40.19}{111.88}{256}
\emline{40.19}{111.88}{257}{41.62}{110.07}{258}
\emline{41.62}{110.07}{259}{43.06}{108.14}{260}
\emline{43.06}{108.14}{261}{44.53}{106.09}{262}
\emline{44.53}{106.09}{263}{46.01}{103.92}{264}
\emline{46.01}{103.92}{265}{47.50}{101.63}{266}
\emline{47.50}{101.63}{267}{49.01}{99.22}{268}
\emline{49.01}{99.22}{269}{50.54}{96.70}{270}
\emline{50.54}{96.70}{271}{52.08}{94.06}{272}
\emline{52.08}{94.06}{273}{54.27}{90.16}{274}
\emline{34.13}{118.47}{275}{35.59}{117.64}{276}
\emline{35.59}{117.64}{277}{37.07}{116.67}{278}
\emline{37.07}{116.67}{279}{38.58}{115.57}{280}
\emline{38.58}{115.57}{281}{40.12}{114.34}{282}
\emline{40.12}{114.34}{283}{41.68}{112.97}{284}
\emline{41.68}{112.97}{285}{43.27}{111.47}{286}
\emline{43.27}{111.47}{287}{44.89}{109.83}{288}
\emline{44.89}{109.83}{289}{46.54}{108.06}{290}
\emline{46.54}{108.06}{291}{48.21}{106.16}{292}
\emline{48.21}{106.16}{293}{49.91}{104.12}{294}
\emline{49.91}{104.12}{295}{51.64}{101.95}{296}
\emline{51.64}{101.95}{297}{53.40}{99.64}{298}
\emline{53.40}{99.64}{299}{55.18}{97.20}{300}
\emline{55.18}{97.20}{301}{58.46}{92.47}{302}
\emline{58.62}{92.48}{303}{60.17}{90.67}{304}
\emline{60.17}{90.67}{305}{61.78}{88.81}{306}
\emline{61.78}{88.81}{307}{63.48}{86.89}{308}
\emline{63.48}{86.89}{309}{65.24}{84.93}{310}
\emline{65.24}{84.93}{311}{67.08}{82.91}{312}
\emline{67.08}{82.91}{313}{68.99}{80.84}{314}
\emline{68.99}{80.84}{315}{70.98}{78.72}{316}
\emline{70.98}{78.72}{317}{73.04}{76.55}{318}
\emline{73.04}{76.55}{319}{75.18}{74.33}{320}
\emline{75.18}{74.33}{321}{77.39}{72.06}{322}
\emline{77.39}{72.06}{323}{79.67}{69.73}{324}
\emline{79.67}{69.73}{325}{84.45}{64.93}{326}
\emline{84.45}{64.93}{327}{89.53}{59.92}{328}
\emline{89.53}{59.92}{329}{94.91}{54.70}{330}
\emline{94.91}{54.70}{331}{103.53}{46.49}{332}
\emline{103.53}{46.49}{333}{112.81}{37.82}{334}
\emline{112.81}{37.82}{335}{126.21}{25.54}{336}
\emline{126.21}{25.54}{337}{132.54}{19.83}{338}
\emline{54.20}{90.17}{339}{56.25}{87.58}{340}
\emline{56.25}{87.58}{341}{58.35}{85.00}{342}
\emline{58.35}{85.00}{343}{60.49}{82.43}{344}
\emline{60.49}{82.43}{345}{62.67}{79.88}{346}
\emline{62.67}{79.88}{347}{64.90}{77.34}{348}
\emline{64.90}{77.34}{349}{67.18}{74.81}{350}
\emline{67.18}{74.81}{351}{69.49}{72.29}{352}
\emline{69.49}{72.29}{353}{71.86}{69.79}{354}
\emline{71.86}{69.79}{355}{74.26}{67.29}{356}
\emline{74.26}{67.29}{357}{76.72}{64.81}{358}
\emline{76.72}{64.81}{359}{79.21}{62.34}{360}
\emline{79.21}{62.34}{361}{81.75}{59.89}{362}
\emline{81.75}{59.89}{363}{84.34}{57.44}{364}
\emline{84.34}{57.44}{365}{86.97}{55.01}{366}
\emline{86.97}{55.01}{367}{89.64}{52.59}{368}
\emline{89.64}{52.59}{369}{92.36}{50.18}{370}
\emline{92.36}{50.18}{371}{95.12}{47.78}{372}
\emline{95.12}{47.78}{373}{97.93}{45.39}{374}
\emline{97.93}{45.39}{375}{100.78}{43.02}{376}
\emline{100.78}{43.02}{377}{103.68}{40.66}{378}
\emline{103.68}{40.66}{379}{106.62}{38.31}{380}
\emline{106.62}{38.31}{381}{109.60}{35.97}{382}
\emline{109.60}{35.97}{383}{112.63}{33.65}{384}
\emline{112.63}{33.65}{385}{115.70}{31.33}{386}
\emline{115.70}{31.33}{387}{118.82}{29.03}{388}
\emline{118.82}{29.03}{389}{121.98}{26.74}{390}
\emline{121.98}{26.74}{391}{125.19}{24.46}{392}
\emline{125.19}{24.46}{393}{128.44}{22.20}{394}
\emline{128.44}{22.20}{395}{131.91}{19.83}{396}
\emline{34.00}{118.33}{397}{35.38}{117.73}{398}
\emline{35.38}{117.73}{399}{36.86}{116.97}{400}
\emline{36.86}{116.97}{401}{38.43}{116.08}{402}
\emline{38.43}{116.08}{403}{40.09}{115.05}{404}
\emline{40.09}{115.05}{405}{41.86}{113.87}{406}
\emline{41.86}{113.87}{407}{43.71}{112.55}{408}
\emline{43.71}{112.55}{409}{45.67}{111.09}{410}
\emline{45.67}{111.09}{411}{47.71}{109.48}{412}
\emline{47.71}{109.48}{413}{49.86}{107.74}{414}
\emline{49.86}{107.74}{415}{52.10}{105.85}{416}
\emline{52.10}{105.85}{417}{54.43}{103.82}{418}
\emline{54.43}{103.82}{419}{56.86}{101.64}{420}
\emline{56.86}{101.64}{421}{59.39}{99.33}{422}
\emline{59.39}{99.33}{423}{62.01}{96.87}{424}
\emline{62.01}{96.87}{425}{64.73}{94.27}{426}
\emline{64.73}{94.27}{427}{67.54}{91.52}{428}
\emline{67.54}{91.52}{429}{70.45}{88.64}{430}
\emline{70.45}{88.64}{431}{73.45}{85.61}{432}
\emline{73.45}{85.61}{433}{76.55}{82.44}{434}
\emline{76.55}{82.44}{435}{79.74}{79.13}{436}
\emline{79.74}{79.13}{437}{83.03}{75.67}{438}
\emline{83.03}{75.67}{439}{86.42}{72.07}{440}
\emline{86.42}{72.07}{441}{89.90}{68.33}{442}
\emline{89.90}{68.33}{443}{93.48}{64.45}{444}
\emline{93.48}{64.45}{445}{97.15}{60.43}{446}
\emline{97.15}{60.43}{447}{104.78}{51.95}{448}
\emline{104.78}{51.95}{449}{112.79}{42.91}{450}
\emline{112.79}{42.91}{451}{121.18}{33.29}{452}
\emline{121.18}{33.29}{453}{132.33}{20.33}{454}
\emline{33.33}{119.00}{455}{35.43}{118.48}{456}
\emline{35.43}{118.48}{457}{37.58}{117.82}{458}
\emline{37.58}{117.82}{459}{39.78}{117.02}{460}
\emline{39.78}{117.02}{461}{42.02}{116.08}{462}
\emline{42.02}{116.08}{463}{44.31}{114.99}{464}
\emline{44.31}{114.99}{465}{46.65}{113.76}{466}
\emline{46.65}{113.76}{467}{49.03}{112.39}{468}
\emline{49.03}{112.39}{469}{51.46}{110.87}{470}
\emline{51.46}{110.87}{471}{53.94}{109.22}{472}
\emline{53.94}{109.22}{473}{56.47}{107.42}{474}
\emline{56.47}{107.42}{475}{59.05}{105.48}{476}
\emline{59.05}{105.48}{477}{61.67}{103.39}{478}
\emline{61.67}{103.39}{479}{64.34}{101.17}{480}
\emline{64.34}{101.17}{481}{67.06}{98.80}{482}
\emline{67.06}{98.80}{483}{69.82}{96.29}{484}
\emline{69.82}{96.29}{485}{72.63}{93.63}{486}
\emline{72.63}{93.63}{487}{75.49}{90.84}{488}
\emline{75.49}{90.84}{489}{78.40}{87.90}{490}
\emline{78.40}{87.90}{491}{81.35}{84.82}{492}
\emline{81.35}{84.82}{493}{84.36}{81.60}{494}
\emline{84.36}{81.60}{495}{87.41}{78.23}{496}
\emline{87.41}{78.23}{497}{90.50}{74.73}{498}
\emline{90.50}{74.73}{499}{93.65}{71.08}{500}
\emline{93.65}{71.08}{501}{96.84}{67.28}{502}
\emline{96.84}{67.28}{503}{100.08}{63.35}{504}
\emline{100.08}{63.35}{505}{103.36}{59.27}{506}
\emline{103.36}{59.27}{507}{106.70}{55.05}{508}
\emline{106.70}{55.05}{509}{110.08}{50.69}{510}
\emline{110.08}{50.69}{511}{113.51}{46.19}{512}
\emline{113.51}{46.19}{513}{116.99}{41.54}{514}
\emline{116.99}{41.54}{515}{120.51}{36.75}{516}
\emline{120.51}{36.75}{517}{124.08}{31.82}{518}
\emline{124.08}{31.82}{519}{127.70}{26.75}{520}
\emline{127.70}{26.75}{521}{132.67}{19.67}{522}
\put(84.67,85.00){\makebox(0,0)[cc]{$1$}}
\put(76.33,85.00){\makebox(0,0)[cc]{$2$}}
\put(68.00,85.00){\makebox(0,0)[cc]{$3$}}
\put(61.33,85.00){\makebox(0,0)[cc]{$4$}}
\end{picture}

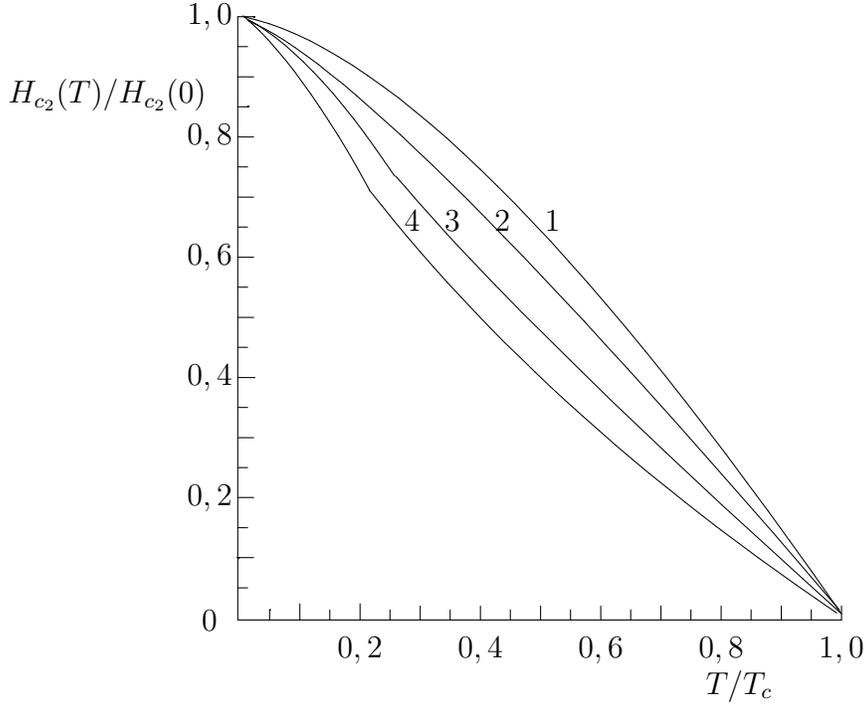
Fig.1.The  temperature dependence $H_{c_2}(T)/H_{c_2}(0)$ at $\lambda_{11} =
0,2$\,,$\lambda_{12} = 0,3$ and values $v_1/ v_2 = 1, 2, 3$ and 4 (curves 1-4
correspondingly).\\
\\

It is easy to see, that with growth of $v_1/v_2$, the curvature
in this dependence changes.  Curves 3 and 4 give the curvature, which was
observed during the experiment in a row of oxidation ceramics. So if there are
heavy carriers in the second band (low speeds on the Fermi surface), the
two-band model qualitatively describes the behavior of $H_{c_2}$ as a function
of temperature in these materials.

Obtained above results let us do the conclusion about qualitative describing
of the experimental data by the behavior of the ratio $H_{c_2}(T)/H_{c_2}
(0)$ as a function of temperature \cite{Canfield} in the intermetallic
compound $MgB_2$.This theory give  also the big values of $H_{c_2}(0)$ in
two-band system in comparison with one-band  case.  For receiving of the
quantitative accordance with the experimental data it is necessary to know the
theory parameters conformably to $MgB_2$, which should be evaluated from the
existing experimental data on the base of two-band model.\\

{\bf Acknowledgment}.\\
\\
I is very grateful to Prof. T.Mishonov  for his interest to our works,
discribend theory of two-band superconductors and for given references about
investigation property of MgB2.

\end{document}